\documentclass{article}

\normalsize
\advance\textheight by \topskip
\twocolumn

\usepackage{times}
\usepackage{graphicx} 
\def\lsi{\raise0.3ex\hbox{$<$\kern-0.75em\raise-1.1ex\hbox{$\sim$}}}
\def\gsi{\raise0.3ex\hbox{$>$\kern-0.75em\raise-1.1ex\hbox{$\sim$}}}

\begin{document}

\title{Propagation of ultrahigh energy cosmic rays and compact sources}
\author{Z. Fodor and S.D. Katz \\
Institute for Theoretical Physics, E\"otv\"os University, \\P\'azm\'any
1, H-1117 Budapest, Hungary}




\maketitle

\begin{abstract}
The clustering of
ultrahigh energy ($>10^{20}$~eV) cosmic rays 
(UHECR) suggests that they might be emitted by compact sources. 
Statistical analysis (Dubovsky et al., 2000)  
estimated the source density.
We extend their analysis to 
give also the confidence intervals (CI) for the source density using
a.) no assumptions on the relationship between clustered and 
unclustered events; b.) nontrivial distributions for the source 
luminosities and energies; c.) the energy dependence of the propagation.
We also determine the probability that a proton created at a distance $r$ with
energy $E$  arrives at earth above a threshold $E_c$. Using this function
one can determine the observed spectrum just by one numerical integration
for any injection spectrum.
The observed 14 UHECR events above $10^{20}$~eV with one doublet gives 
for the source densities $180_{-165}^{+2730}\cdot 10^{-3}$~Mpc$^{-3}$ 
(on the 68\% confidence level). 
\end{abstract}

\section{INTRODUCTION}

The interaction of protons with photons of the cosmic microwave 
background predicts a sharp drop in the cosmic ray flux above
the GZK cutoff around $5\cdot 10^{19}$~eV 
(Greisen, 1966; Zatsepin and Kuzmin, 1966). The 
available data shows no such drop. 
About 20 events above $10^{20}$~eV
were observed by a number of experiments such as AGASA 
(Takeda et al., 1998), Fly's Eye (Bird et al., 1993),
Haverah Park (Lawrence et al., 1991),
Yakutsk (Efimov et al., 1991) and HiRes (Kieda et al., 1999).
Future experiments, particularly Pierre Auger (Boratav, 1996;
Guerard, 1999; Bertou et al. 2000), will have a much higher statistics.
Since above
the GZK energy the attenuation length of particles is a few tens
of megaparsecs 
(Yoshida and Teshima, 1993 ; Aharonian and Cronin, 1994;
Protheroe and  Johnson, 1996; Bhattacharjee and  Sigl, 2000;
Achterberg et al., 1999; T. Stanev et al., 2000)
if an ultrahigh energy cosmic ray (UHECR) is
observed on earth it is usually assumed that it is 
produced in our vicinity. 

At these high energies the galactic and extragalactic magnetic 
fields do not affect the orbit of the cosmic rays, thus they 
should point back to their origin within a few degrees. In 
contrast to the low energy cosmic rays one can use UHECRs
for point-source search astronomy. 
Though there are some peculiar clustered events, which we discuss in 
detail,  the overall distribution of UHECRs on the sky is practically isotropic. 
This observation is rather surprising since in principle only a few 
astrophysical sites are capable to accelerate such particles. 

There are several ways to look for
the source inhomogeneity from the energy spectrum and 
spatial directions of UHECRs. One possibility is to 
assume that the source density of UHECRs is proportional to 
the galaxy densities (Waxman et al., 1997; Giller et al., 1980;
Hill and Schramm, 1985); or one can analyze 
the clustering of the unknown sources by some 
correlation length (Bahcall and Waxman 2000).

Clearly, the arrival directions of 
the UHECRs measured by experiments show some peculiar clustering: 
some events are grouped within $ \sim 3^o$, the typical angular 
resolution of an experiment. Above $4\cdot 10^{19}$ eV 92 cosmic ray events 
were detected, including 7 doublets and 2 triplets. 
Above $10^{20}$ eV one doublet out of 14 events were found 
(Uchihori et al., 2000). 
The chance probability of such a clustering from uniform distribution is 
rather small (Uchihori et al., 2000; Hayashida et al., 1996; Tinyakov
and Tkachev 2001a, Tinyakov and Tkachev 2001b,). 

The clustered features of the events initiated
an interesting statistical analysis  
assuming compact UHECR sources (Dubovsky et al., 2000). The authors found
a large number, $\sim 400$ for the number of 
sources inside a GZK sphere of 25~Mpc. 
We extend their analysis to
give also the CIs for the source density using
a.) no assumptions on the relationship between clustered and
unclustered events; b.) nontrivial distributions for the source
luminosities and energies; c.) the energy dependence of the propagation.
We also determine the probability that a proton created at a distance $r$ with
energy $E$  arrives at earth above a threshold $E_c$.

As we show the most probable value
for the source density is really large; however, the 
statistical significance of this result is rather weak. At
present the small number of UHECR events allows a 95\%
CI for the source density which spreads 
over four orders of magnitude. Since future experiments,
particularly Pierre Auger, will have a much
higher significance on clustering (the expected 
number of events of $10^{20}$ eV and above is 60 per year, 
we present give results also 
for larger number of UHECRs above $10^{20}$ eV.

In order to avoid the assumptions of (Dubovsky et al., 2000) 
a combined analytical 
and Monte-Carlo technique will be presented adopting the 
conventional picture of protons as the ultrahigh energy cosmic rays.
Our analytical approach of Section \ref{sec_anal}
gives the event clustering 
probabilities for any space, luminosity and energy distribution  
by using a single additional function $P(r,E;E_c)$, the probability
that a proton created at a distance $r$ with energy $E$ arrives
at earth above the threshold energy $E_c$ (Bahcall and Waxman, 2000). 
With our Monte-Carlo technique of Section \ref{sec_monte} we
determine the probability function $P(r,E;E_c)$ for 
a wide range of parameters. Our results 
for the present and future UHECR statistics are presented in Section 
\ref{sec_res}. 
We summarize in Section \ref{sec_sum}.  

\section{ANALYTICAL APPROACH} \label{sec_anal}

The number of UHECRs emitted by a source of $\lambda$ luminosity
during a period $T$ follows the Poisson distribution. 
However, not all
emitted UHECRs will be detected. They might loose their energy during
propagation or can simply go to the wrong direction.
For UHECRs the energy loss
is dominated by the pion production in interaction with
the cosmic microwave background radiation. In a recent analysis 
(Bahcall and Waxman, 2000) 
the probability function $P(r,E,E_c)$ was presented
for three specific threshold energies. This function gives 
the probability that a proton
created at a given distance from earth (r) with some energy (E) is detected
at earth above some energy threshold ($E_c$). 

The features of the Poisson distribution enforce us to take
into account the fact that the sky is not
isotropically observed. 

The probability of detecting $k$ events from a source at distance 
$r$ with energy $E$ can be obtained by simply including the factor
$P(r,E,E_c) A\eta/(4\pi r^2)$ in the Poisson distribution:
\begin{eqnarray}
p_k({\bf x},E,j)
=\frac{\exp\left[  -P(r,E,E_c)\eta j/r^2 \right] }{k!}\times \nonumber\\
\left[ P(r,E,E_c)\eta j/r^2\right] ^k, \label{poiss2}
\end{eqnarray}
where we introduced $j=\lambda T A/(4\pi)$ and $A\eta/(4\pi r^2)$ 
is the 
probability that an emitted UHECR points to a detector of 
area $A$. The factor $\eta$ represents the visibility of the source,
which was determined by spherical astronomy. 
We denote the space, energy and 
luminosity  distributions of the sources by $\rho({\bf x})$, 
$c(E)$ and $h(j)$, respectively. The probability of detecting $k$
events above the threshold $E_c$ from a single source 
randomly positioned within a sphere of radius $R$ is
\begin{eqnarray}\label{P_k}
P_k=\int_{S_R} dV\; \rho({\bf x}) \int_{E_c}^{\infty} 
dE\; c(E) \int_0^{\infty} dj\; h(j) \times \nonumber \\ 
\frac{\exp\left[ -P(r,E,E_c)\eta j/r^2\right] }{k!} \left[
P(r,E,E_c)\eta j/r^2 \right] ^k. \end{eqnarray}

Denote the total number of sources within the sphere of 
sufficiently large radius (e.g. several times the GZK radius)
by $N$ and the number of sources that gave $k$ detected events by
$N_k$. Clearly, $N=\sum_0^{\infty}N_i$ and the total number of detected
events is $N_e=\sum_0^{\infty}i N_i$. The probability that for $N$
sources the number of different detected multiplets are $N_k$ is:
\begin{equation}\label{distribution}
P(N,\{N_k\})=N!\prod_{k=0}^{\infty} \frac{1}{N_k!}P_k^{N_k}.
\end{equation}
For a given set of 
unclustered and clustered events ($N_1$ and 
$N_2,N_3$,...) 
inverting the $P(N,\{N_k\})$ distribution function 
gives the most probable value for the number 
of sources and also the CI for it.

Note, that $P_k$ and then $P(N,\{N_k\})$ are easily determined by
a well behaving four-dimensional numerical integration
for any $c(E)$, $h(j)$ and $\rho (r)$ distribution functions. 
In order to illustrate the uncertainties and sensitivities of the
results we used a few different 
choices for these distribution functions. 

For $c(E)$ we studied three possibilities. The most 
straightforward choice is the extrapolation of the `conventional
high energy component' $\propto E^{-2}$. Another possibility is
to use a stronger
fall-off of the spectrum at energies just below the GZK cutoff,
e.g. $\propto E^{-3}$. 
The third possibility is to assume 
that UHECRs are some decay products of metastable superheavy
particles (Berezinsky et al., 1997; Kuzmin and Rubakov, 1998; 
Birkel and Sarkar, 1998; Sarkar 2000; Fodor and Katz 2001b).
According to (Birkel and Sarkar, 1998)
the superheavy particles decay into quarks and gluons which initiate 
multi-hadron cascades through gluon bremstrahlung. 

In the recent analysis (Dubovsky et al., 2000)  
the authors have shown that for a fixed
set of multiplets the minimal density of sources can be obtained 
by assuming a delta-function distribution for $h(j)$. We 
studied both this limiting luminosity: $h(j)=\delta(j-j_*)$ 
and a more realistic one
with Schechter's luminosity function (Schechter 1976), which
can be given as:
$h(j)dj=h\cdot (j/j_*)^{-1.25}\exp(-j/j_*)d(j/j_*)$. 

The space distribution of sources can be given based on some
particular survey of the distribution of nearby galaxies 
or on a correlation length $r_0$ characterizing 
the clustering features of sources. For simplicity 
the present analysis deals with a homogeneous distribution of
sources.

\section{MONTE-CARLO STUDY OF THE PROPAGATION} \label{sec_monte}

In our Monte-Carlo
approach we determined the propagation of UHECR protons 
on an event by event 
basis. The inelasticity of Bethe-Heitler
pair production is small
($\approx 10^{-3}$), thus we used a continuous energy loss approximation for
this process. The inelasticity of pion-photoproduction is larger
($\approx 0.2 -0.5$) in the energy range of interest, thus there are only a
few tens of such interactions during the propagation. Due to the Poisson
statistics and the spread of the
inelasticity, we will see a spread in the energy spectrum even if the
injected spectrum is mono-energetic.

In our simulation protons are propagated in small steps
($10$~kpc), and after each step the energy losses due to pair
production, pion production and the adiabatic expansion are calculated.
During the simulation we keep track of the current energy of the proton
and its total displacement. 
For the proton interaction lengths and inelasticities
we used the values of (Bhattacharjee and  Sigl, 2000;
Achterberg et al., 1999). The deflection 
due to magnetic field is not taken into account
(for a recent Monte-Carlo on it
see eg. Stanev et al., 2000).

\begin{figure}
\vspace*{2.0mm}
\includegraphics[width=8.3cm]{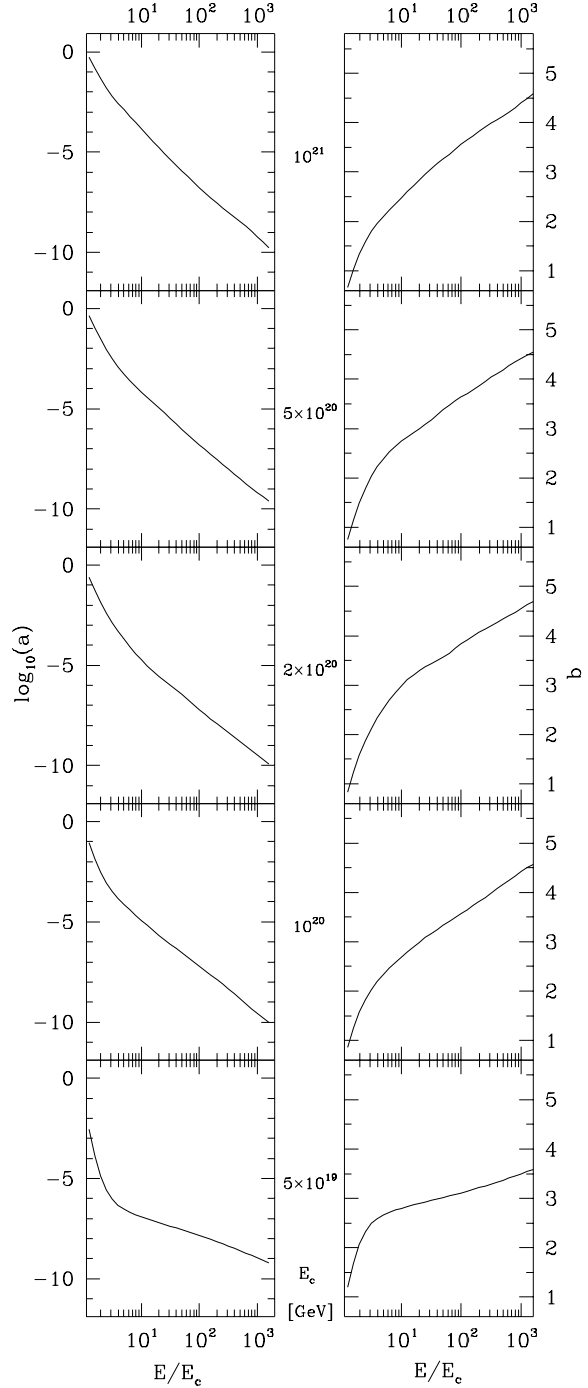}
\caption{\label{gzk}
{ \sl The functions $a(E/E_c)$ --left panel-- and $b(E/E_c)$ 
--right panel-- for the probability distribution function 
$P(r,E,E_c)$ using the parametrization 
$\exp[-a\cdot(r/1\ {\rm Mpc})^b]$ for five different threshold
energies ($5\cdot 10^{19}$~eV, $10^{20}$~eV, $2\cdot 10^{20}$~eV,
$5\cdot 10^{20}$~eV and $10^{21}$~eV).
}}
\end{figure}

To
cover a broad energy range we  used the parametrization
\begin{equation} \label{parametr}
P(r,E,E_c)=\exp\left[ -a\cdot(r/1\ {\rm Mpc})^b\right].
\end{equation}
Fig. \ref{gzk} shows the functions $a(E/E_c)$ and $b(E/E_c)$
for a range of three orders of magnitude and for five different
threshold energies. Just using the functions of $a(E/E_c)$ and 
$b(E/E_c)$, thus a parametrization of $P(r,E,E_c)$ one can obtain the 
observed energy spectrum for any injection spectrum without additional
Monte-Carlo simulation.

\section{RESULTS}\label{sec_res}

In order to determine the CIs for the  
source densities we used the frequentist method (Groom et al., 2000).
We wish to set limits
on S, the source density. Using our Monte-Carlo based
$P(r,E,E_c)$ functions and our analytical technique we
determined $p(N_1,N_2,N_3,...;S;j_*)$, which gives the probability of 
observing $N_1$ singlet, $N_2$ doublet, $N_3$ 
triplet etc. events
if the true value of the density is $S$ and the central value of
luminosity is $j_*$. 
For a given set of 
$\{N_i,i=1,2,...\}$ the above probability distribution as a 
function of $S$ and $j_*$ determines the 68\% and 95\%
confidence level regions in the $S-j_*$ plane.
For different choices of $c(E)$
and $h(j)$ see Table \ref{results} for the 
calculated densities. 
Our results for the Dirac-delta luminosity
distribution are in agreement with
the result of (Dubovsky et al., 2000) 
within the error bars. Neverthless, there is a
very important message.
The CIs are are so large that on the 95\%  
confidence level two orders of magnitude smaller densities than
suggested as a lower bound by (Dubovsky et al., 2000) are also possible.

It is of particular interest to study higher statistics 
too, and determine CIs for these cases. 
We performed a detailed analysis on this question 
(Fodor and Katz 2001a).  
An interesting feature of the  results is that ''doubling'' the present
statistics with the same clustering features (in the case studied by 
the table this means one new doublet out of 10 new events) reduces
the CIs by an order of magnitude. 
The study of even higher statistics leads to the conclusion 
that experiments in the near future
with approximately 200 UHECR events can tell at least the order of 
magnitude of the source density. 

\begin{table}[htb]
\begin{center}\begin{tabular}{c|c|c}
$c(E)$ & $h(j)$ & {\bf 14 events 1 doublet} \\
\hline $\propto E^{-2}$ & $\propto \delta$ &
$      2.77 _{-       2.53 (       2.70 )}^{+      96.1  (     916 )}  $\\
\hline $\propto E^{-2}$ & $\propto$ SLF &
$       36.6 _{-        34.3 ( 35.9)}^{+844(4268)}$ \\
\hline $\propto E^{-3}$ & $\propto \delta$ &
$      5.37 _{-       4.98 (       5.25 )}^{+      80.2  (     624 )}  $\\
\hline $\propto E^{-3}$ & $\propto$ SLF &
$       180 _{- 165 ( 174)}^{+2730(8817)}$      \\
\hline $\propto$ decay  & $\propto \delta$ &
$      3.61 _{-       3.30 (       3.51 )}^{+     116    (    1060 )}  $\\
\hline $\propto$ decay  & $\propto$ SLF &
$       40.9 _{-        38.3 ( 40.1)}^{+856(4345)}$ \\
\end{tabular}
\vspace{0.3cm}
\caption{\label{results}
{ \sl The most probable values for the source densities
and their error bars given by the 68\% and 95\% confidence
level regions (the latter in parenthesis). 
The numbers are in units of $10^{-3}$~Mpc$^{-3}$
The three possible energy spectrums 
are given by a distribution proportional to $E^{-2}$, $E^{-3}$, 
or by the decay of a $10^{12}$ GeV particle (denoted by
``decay''). The luminosity distribution can be proportional 
to a Dirac-delta or to Schechter's luminosity function 
(denoted by ``SLF'').
}}
\end{center}\end{table}

\section{SUMMARY} \label{sec_sum}

We presented a technique in order to statistically analyze the
clustering features of UHECR. The technique can be applied
for any model of UHECR assuming small deflection. The key role 
of the analysis is played by the $P_k$ functions defined
by eqn. (\ref{P_k}), which is the probability of detecting $k$
events above the threshold from a single source. Using a 
combinatorial expression of eqn. (\ref{distribution}) the 
probability distribution for any set of multiplets can be given
as a function of the source density.

We discussed several types of energy and luminosity distributions
for the sources and gave the most probable source densities
with their CIs for present and future 
experiments. 

The probability $P(r,E,E_c)$ that a proton created at a distance $r$
with energy $E$ arrives above the threshold $E_c$ 
(Bahcall and Waxman, 2000) is
determined and parametrized for a wide range of threshold energies. 
This result can be used to obtain the observed energy spectrum of the UHECR
for arbitrary injection spectrum. 

In ref. (Dubovsky et al., 2000) the authors analyzed the statistical features
of clustering of UHECR, which provided constraints on astrophysical 
models of UHECR if the number of clusters is small, by giving a 
bound from below. In our paper we have shown that there is some 
constraint, but it is far from being tight. At present statistics 
the 95\% CIs usually span 4 orders of magnitude. 
Two orders of magnitude smaller numbers than the prediction
of (Dubovsky et al., 2000) (their eqn. (13) suggests for the density 
of sources $\sim 6\cdot 10^{-3}$~Mpc$^{-3}$) can also be obtained.  
Adding 10 new events with an additional doublet 
the CI can be reduced to 3 orders of
magnitude and the increase of the UHECR events to 200 can tell
at least the order of magnitude of the source density. 

More details of the present analysis can be found in
(Fodor and Katz, 2001a). Note, that a similar study
based on the Z-burst scenario (Fargion et al., 1999; Weiler,
1999) can be carried out which gives the mass of the heaviest neutrino
(Fodor et al., 2001).

\section{ACKNOWLEDGEMENTS}

We thank K. Petrovay for clarifying some issues in spherical astronomy.
This work was partially supported by 
Hungarian Science Foundation
grants No. OTKA-T29803/\-T22929-FKP-0128/\-1997/\-OMMU708/\-IKTA/\-NIIF.

\vspace*{0.5cm}\noindent
{\Large {\bf References}}{\small
\begin{description}
\item Achterberg,A. et al., astro-ph/9907060.
\item Aharonian,F.A. and Cronin,J.W., Phys. Rev. {\bf D50}, 1892 (1994).
\item Bahcall,J.N., and Waxman,E., Astrophys.J. {\bf 541}, 707 (2000).
\item Berezinsky,V. and Kachelrie{\ss},M., Phys.Lett. {\bf B434}, 61 (1998).
\item Berezinsky,V., Kachelrie{\ss},M.,   and Vilenkin,A.,
Phys. Rev. Lett. {\bf 79}, 4302 (1997).
\item Bertou,X., Boratav,M., and Letessier-Selvon,A., astro-ph/0001516.
\item Bhattacharjee,P., and Sigl,G., Phys. Rep. {\bf 327}, 109 (2000).
\item Bird,D.J. et al., Phys. Rev. Lett. {\bf 71}, 3401 (1993);
Astrophys J. {\bf 424}, 491 (1994); ibid {\bf 441}, 144 (1995).
\item Birkel,M. and Sarkar,S., Astropart. Phys. {\bf 9}, 297 (1998).
\item Boratav,M.,  Nucl. Phys. Proc. {\bf 48}, 488 (1996).
\item Dawson,B.R., Meyhandan,R., and Simpson,K.M.,
Astropart. Phys. {\bf 9}, 331 (1998).
\item Dubovsky,S.L., Tinyakov,P.G., and Tkachev,I.I.,
Phys. Rev. Lett. 85 (2000) 1154.
\item Efimov,N.N. et al., "Proc. Astrophysical
Aspects of the Most Energetic Cosmic Rays", p. 20, eds.
M.~Nagano and F.~Takahara, World Sci., Singapore, 1991.
\item Fargion,D., Mele,B., and Salis,A., Astrophys. J. 517 (1999) 725.
\item Fodor,Z. and  Katz,S.D., Phys. Rev. {\bf D63}, 023002 (2001).
\item Fodor,Z. and  Katz,S.D., Phys. Rev. Lett. {\bf 86}, 3224 (2001).
\item Fodor,Z., Katz.S.D., and Ringwald,A., hep-ph/0105064.
\item Giller,M., Wdowczyk,J., and Wolfendale,A., J. Phys. {\bf G6}, 1561 (1980).
\item Greisen,K., Phys.Rev.Lett. {\bf 16}, 748 (1966).
\item Groom,D.E. et al., Eur. Phys. J. {\bf C15}, 1 (2000).
\item Guerard,C.K., ibid {\bf 75A}, 380 (1999).
\item Hayashida,N. et al., Phys. Rev. Lett. {\bf 77}, 1000 (1996).
\item Hill,C.T., and Schramm,D.N., Phys. Rev. {\bf D31}, 564 (1985).
\item Kieda,D. et al., to appear in Proc. of the 26th ICRC, Salt Lake,
\item Kuzmin,V.A. and Rubakov,V.A., Phys. Atom. Nucl. {\bf 61}, 1028 (1998).
\item Lawrence,M.A., Reid,R.J.O., and Watson,A.A.,
J. Phys. {\bf G17}, 773 (1991).
\item Protheroe,R.J., Johnson,P., Astropart. Phys. {\bf 4}, 253 (1996).
\item Sarkar,S., hep-ph/0005256.
\item Stanev,T. et al., astro-ph/0003484.
\item Takeda,M. et al., Phys. Rev. Lett. {\bf 81}, 1163 (1998); astro-ph/\-9902239.
\item Tinyakov,P.G., and Tkachev,I.I., astro-ph/0102101.
\item Tinyakov,P.G., and Tkachev,I.I., astro-ph/0102476. 
\item Uchihori,Y. et al., Astropart. Phys. {\bf 13}, 151 (2000).
\item Yoshida,S., Teshima,M., Prog. Theor. Phys. {\bf 89}, 833 (1993).
\item Waxman,E., Fisher,K.B., and Piran,T., Astrophys. J. {\bf 483}, 1 (1997).
\item Weiler,T.J., Astropart. Phys. {\bf 11} (1999) 303; Astropart. Phys. {\bf 12}
(2000) 379 (Erratum).
\item Zatsepin,G.T. and Kuzmin,V.A., Pisma Zh.Exp.Teor.Fiz. {\bf 4}, 114 (1966).
\end{description}}
\end{document}